\documentclass[sigplan]{acmart} 
\renewcommand\footnotetextcopyrightpermission[1]{}

\acmSubmissionID{\#28}

\usepackage{booktabs} 
\usepackage{ifthen}
\usepackage[inline]{enumitem}
\usepackage{prp-macros}
\usepackage{subcaption}
\usepackage{fancyhdr}
\usepackage{framed}
\usepackage{hyperref}
\usepackage{tabularx}
\usepackage{xcolor,colortbl}
\usepackage[printonlyused]{acronym}
\usepackage{cleveref} 
\usepackage{todonotes}
\usepackage[percent]{overpic}
\usepackage[subtle]{savetrees}

\crefformat{section}{\S#2#1#3}
\crefformat{subsection}{\S#2#1#3}
\crefformat{subsubsection}{\S#2#1#3}
\crefmultiformat{section}{\S#2#1#3}{ and~\S#2#1#3}{, \S#2#1#3}{, and~\S#2#1#3}
\crefformat{figure}{Fig.\nobreak\hspace{0.25em}#2#1#3}
\crefformat{table}{Table\nobreak\hspace{0.25em}#2#1#3}
\crefmultiformat{figure}{Figure #2#1#3}{ and~Figure #2#1#3}{, Figure #2#1#3}{, and~Figure#2#1#3}
\crefname{section}{\S}{\S\S}
\crefname{subsection}{\S}{\S\S}
\crefname{subsubsection}{\S}{\S\S}
\crefname{table}{Table}{Tables}
\crefname{figure}{Fig.}{Figs.}

\newlist{inlinelist}{enumerate*}{1}
\setlist*[inlinelist,1]{%
  label=(\roman*),
}

\newcommand{\disable}[1]{} 

\newcommand{\setA}{\texttt{A}}
\newcommand{\setB}{\texttt{B}}
\newcommand{\setC}{\texttt{C}}
\newcommand{\setD}{\texttt{D}}
\newcommand{\setE}{\texttt{E}}
\newcommand{\setF}{\texttt{F}}

\newcommand{\magic}{\textsc{Hector}\xspace}
\newcommand{\trustjs}{\textsc{ownWorkshopPaper}\xspace}
\newcommand{\deanon}{}

\renewcommand{\trustjs}{TrustJS\xspace}
\renewcommand{\deanon}{\StopCensoring}

\definecolor{enclavegreen}{RGB}{104,188,54}

\newcommand{\webrtc}{\ac{WebRTC}\xspace}

\newboolean{showcomments}
\setboolean{showcomments}{true}
\ifthenelse{\boolean{showcomments}}
{ \newcommand{\mynote}[3]{
	\textcolor{#3}{$\blacktriangleright${\bfseries\sffamily\scriptsize#1: }\small#2$\blacktriangleleft$}
}}
{ \newcommand{\mynote}[3]{}}

\definecolor{c1}{RGB}{215,48,39}
\definecolor{c2}{RGB}{252,141,89}
\definecolor{c3}{RGB}{254,224,144}
\definecolor{c4}{RGB}{255,225,191}
\definecolor{c5}{RGB}{224,243,248}
\definecolor{c6}{RGB}{145,191,219}
\definecolor{c7}{RGB}{69,117,180}

\startPage{1}

\setcopyright{none}

\usepackage{booktabs}   
\usepackage{subcaption} 
\usepackage{balance}
\usepackage{censor}

\begin{document}
\deanon
\title[\magic: Using Untrusted Browsers to Provision Web Applications]{\magic: Using Untrusted Browsers\\to Provision Web Applications}

\author{David Goltzsche}
\affiliation{
  \institution{TU Braunschweig, Germany}
}
\email{goltzsche@ibr.cs.tu-bs.de}

\author{Tim Siebels}
\affiliation{
  \institution{TU Braunschweig, Germany}
}
\email{siebels@ibr.cs.tu-bs.de}

\author{Lennard Golsch}
\affiliation{
  \institution{TU Braunschweig, Germany}
}
\email{golsch@ibr.cs.tu-bs.de}

\author{R{\"u}diger Kapitza}
\affiliation{
  \institution{TU Braunschweig, Germany}
}
\email{rrkapitz@ibr.cs.tu-bs.de}

\renewcommand{\shortauthors}{D. Goltzsche, T. Siebels, L. Golsch and R. Kapitza}


\begin{abstract}


Web applications are on the rise and rapidly evolve into more and more mature replacements for their native counterparts.
This disruptive trend is mainly driven by the attainment of platform-independence and instant deployability.
On top of this, web browsers offer the opportunity for seamless browser-to-browser communication for distributed interaction.

In this paper, we present \magic, a novel web application framework that transforms web browsers into a distributed application-centric computing platform.
\magic enables offloading application logic to users, thereby improving user experience with lower latencies while generating less costs for service providers.
Following the programming paradigm of Function-as-a-Service, applications are decomposed into functions so they can be managed efficiently and deployed in a responsive, scalable and lightweight fashion.
In case of client-side resource shortage or unresponsive clients, execution falls back to a traditional cloud-based infrastructure.
\magic combines WebAssembly for multi-language computations at near-native speed, WebRTC for browser-to-browser communication and trusted execution as provided by the Intel Software Guard Extensions so browsers can trust each other's computations.
We evaluate \magic by implementing a digital assistant as well as a recommendation system.
Our evaluation shows that \magic achieves lower end-user latencies while generating less costs than traditional deployments. 
Additionally, we show that \magic scales linearly with increasing client numbers and can cope well with unresponsive clients.

\end{abstract}

\maketitle

\acrodef{SPA}{single-page application}
\acrodefplural{SPA}{single-page applications}
\acrodef{EPC}{enclave page cache}
\acrodef{EPCM}{enclave page cache map}
\acrodef{MMU}{Memory Management Unit}
\acrodef{AEX}{Asynchronous Enclave Exit}
\acrodef{AEP}{Asynchronous Exit Pointer}
\acrodef{ISR}{Interrupt Service Routine}
\acrodef{OS}{operating system}
\acrodef{DoS}{Denial-of-Service}
\acrodef{DDoS}{Distributed Denial-of-Service}
\acrodef{SDK}{software development kit}
\acrodef{TCB}{trusted computing base}
\acrodef{TOCTTOU}{time-of-check-to-time-of-use}
\acrodef{EDL}{Enclave Description Language}
\acrodef{API}{application programming interface}
\acrodef{ASLR}{address space layout randomisation}
\acrodef{IP}{intellectual property}
\acrodef{JIT}{just-in-time}
\acrodef{EDL}{enclave description language}
\acrodef{QE}{Quoting Enclave}
\acrodef{IAS}{Intel Attestation Service}
\acrodef{SPID}{Service Provider ID}
\acrodef{RIA}{Rich Internet Application}
\acrodef{QoS}{Quality of Service}
\acrodef{HMAC}{keyed-Hash Message Authentication Code}
\acrodef{TEE}{trusted execution environment}
\acrodef{VPN}{virtual private network}
\acrodef{MEE}{Memory Encryption Engine}
\acrodef{ISP}{Internet service provider}
\acrodefplural{ISP}{Internet service providers}
\acrodef{DLP}{data leak prevention}
\acrodef{MTU}{maximum transmission unit}
\acrodef{RTT}{round-trip time}
\acrodef{AWS}{Amazon Web Services}
\acrodef{EC2}{Elastic Compute Cloud}
\acrodef{IDPS}{intrusion detection and prevention system}
\acrodef{DPI}{deep packet inspection}
\acrodef{TPM}{Trusted Platform Module}
\acrodef{IETF}{Internet Engineering Task Force}
\acrodef{IPC}{inter-process communication}
\acrodef{HPKP}{HTTP public key pinning}
\acrodef{CDF}{cumulative distribution function}
\acrodef{MAC}{message authentication code}
\acrodefplural{LOC}{lines of code}
\acrodef{MITM}{man-in-the-middle}
\acrodef{LKL}{Linux Kernel Library}
\acrodef{NaCl}{Native Client}
\acrodef{PNaCl}{Portable Native Client}
\acrodef{JVM}{Java Virtual Machine}
\acrodef{FaaS}{Function-as-a-Service}
\acrodef{ME}[Intel ME]{Intel Management Engine}
\acrodef{WABT}{WebAssembly Binary Toolkit}
\acrodef{RSS}{resident set size}
\acrodef{CPU}{Central Processing Unit}
\acrodef{VM}{virtual machine}
\acrodef{CA}{Certificate Authority}
\acrodef{PWA}{progressive web application}
\acrodefplural{PWA}{progressive web applications}

\acrodef{API}{Application Programming Interface}
\acrodef{AEX}{Asynchronous Enclave Exit}
\acrodef{AWS}{Amazon Web Services}

\acrodef{BIOS}{Basic Input/Output System}
\acrodef{BOINC}{Berkeley Open Infrastructure for Network Computing}
\acrodef{CDN}{Content Delivery Network}
\acrodef{CPU}{Central Processing Unit}
\acrodef{CSS}{Cascading Style Sheets}

\acrodef{DMA}{Direct Memory Access}
\acrodef{DOM}{Document Object Model}
\acrodef{DoS}{Denial of Service}
\acrodef{DTLS}{Datagram Transport Layer Security}

\acrodef{ECDH}{Elliptic Curve Diffie-Hellman key exchange}
\acrodef{EDL}{Enclave Definition Language}
\acrodef{EPC}{Enclave Page Cache}
\acrodef{EPCM}{Enclave Page Cache Map}
\acrodef{EPID}{Enhanced Privacy ID}
\acrodef{FaaS}{Function-as-a-Service}
\acrodef{FTP}{File Transfer Protocol}
\acrodef{GDB}{GNU Debugger}
\acrodef{HTML}{Hypertext Markup Language}
\acrodef{HTTP}{Hypertext Transfer Protocol}
\acrodef{IaaS}{Infrastructure-as-a-Service}
\acrodef{IAS}{Intel Attestation Service}
\acrodef{ICE}{Interactive Connectivity Establishment}
\acrodef{ICMP}{Internet Control Message Protocol}
\acrodef{IO}{Input/Output}
\acrodef{IoT}{Internet of Things}
\acrodef{IP}{Internet Protocol}
\acrodef{IR}{Intermediate Representation}

\acrodef{JIT}{Just In Time}
\acrodef{JSON}{JavaScript Object Notation}
\acrodef{JVM}{Java Virtual Machine}

\acrodef{MAC}{Message Authentication Code}
\acrodef{MitM}{Man in the Middle}
\acrodef{NAT}{Network Address Translation Protocol}
\acrodef{NPM}{NodeJS Package Manager}
\acrodef{NLU}{natural-language understanding}

\acrodef{PaaS}{Platform-as-a-Service}
\acrodef{PRM}{Processor Reserved Memory}
\acrodef{QE}{Quoting Enclave}
\acrodef{REST}{Representational State Transfer}
\acrodef{SaaS}{Software-as-a-Service}
\acrodef{SECS}{SGX Enclave Control Structure}
\acrodef{SDK}{Software Development Kit}
\acrodef{SDP}{Session Description Protocol}
\acrodef{SGX}{Software Guard Extensions}
\acrodef{SMM}{System Management Mode}
\acrodef{SPA}{Single Page Application}
\acrodef{SRTP}{Secure Real Time Transport Protocol}
\acrodef{SSA}{State Save Area}
\acrodef{STUN}{Session Traversal Utilities for NAT}
\acrodef{SVN}{Security Version Number}

\acrodef{TCB}{Trusted Computing Base}
\acrodef{TCP}{Transmission Control Protocol}
\acrodef{TCS}{Thread Control Structure}
\acrodef{TEE}{Trusted Execution Environment}
\acrodef{TLS}{Transport Layer Security}
\acrodef{TPM}{Trusted Platform Module}
\acrodef{TURN}{Traversal Using Relays around NAT}
\acrodef{TRE}{Trusted Remote Entity}
\acrodef{TXT}{Trusted Execution Technology}

\acrodef{UUID}{Universally Unique Identifier}
\acrodef{URL}{Uniform Resource Locator}
\acrodef{W3C}{World Wide Web Consortium}
\acrodef{WebRTC}{Web Real-Time Communication}
\acrodef{WWW}{World Wide Web}

\DeclareRobustCommand{\compicon}{\raisebox{-0.2em}{\includegraphics[width=1em,page=8]{img/overview.pdf}}}
\begin{figure*}
\begin{subfigure}{0.9\columnwidth}
	\includegraphics[width=\linewidth,page=2]{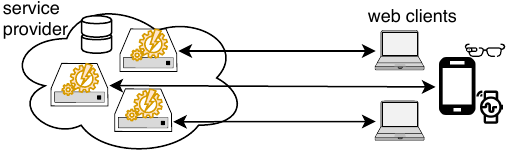}
	\caption{cloud computing}\label{fig:webapp-arch:a}
\end{subfigure}
\begin{subfigure}{0.9\columnwidth}
	\includegraphics[width=\linewidth,page=4]{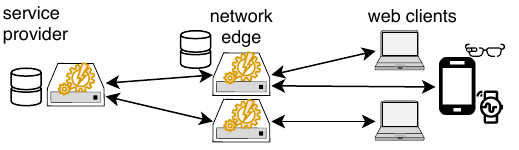}
	\caption{edge computing}\label{fig:webapp-arch:b}
\end{subfigure}
\\
\vspace{0.75cm}
\begin{subfigure}{0.9\columnwidth}
	\includegraphics[width=\linewidth,page=5]{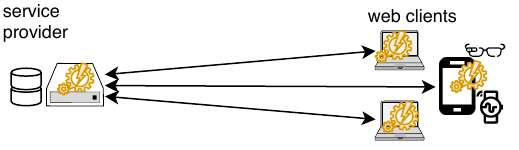}
	\caption{traditional offloading}\label{fig:webapp-arch:c}
\end{subfigure}
\begin{subfigure}{0.9\columnwidth}
	\includegraphics[width=\linewidth,page=7]{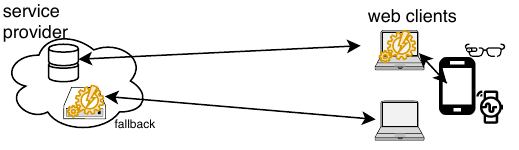}
	\caption{offloading with \magic}
	\label{fig:webapp-arch:d}
\end{subfigure}
\vspace{0.5cm}
\caption{Different approaches for improving scalability and responsiveness of web applications by moving computations (\compicon) to different locations: scaling based on cloud infrastructures (a), shorter latencies and reduced bandwidth by edge computing (b), limited traditional offloading of computation (c); and offloading of computations to untrusted clients with \magic, while maintaining a centralised fallback infrastructure (d).}
\label{fig:webapp-arch}
\end{figure*}
\section{Introduction} \label{sec:intro}
Web applications are about to become the de facto standard for deploying software across various platforms \eg desktop computers, laptops and mobile phones.
So-called \acp{PWA} even create user experiences similar to native ones by offering more functionality, such as offline usage or push notifications while liberating users from installing native apps.
In fact, 50\% of smartphone users favour such \acp{PWA} to avoid installing apps~\cite{google-web-vs-app}.
Popular \acp{PWA} are offered by Spotify, Uber or Tinder~\cite{pwas}.

Reasons for this trend are 
\begin{inlinelist}
	\item improved platform-independence through standardised browser APIs;
	\item the possibility of instant deployment; and
	\item high acceptance of end-users.
\end{inlinelist}
To be widely used, web applications need to scale well to cope with peak demands, and need to provide good responsiveness, especially when interacting with users.
Applications that do not fulfil these properties suffer from bad user experience, resulting in a dissatisfied and, therefore, shrinking user base.

Orthogonal approaches have been established to improve scalability and responsiveness of web applications.
First, scalability on server-side can be improved by deploying applications on highly scalable cloud-based architectures such as \ac{AWS}~\cite{aws} (\Cref{fig:webapp-arch:a}).
Second, edge computing~\cite{shi2016edge,satyanarayanan2017emergence} or \acp{CDN}~\cite{dilley2002globally,nygren2010akamai} can be used to move computations or contents closer to end-users, which improves network latencies and, therefore, the responsiveness of the whole application (\Cref{fig:webapp-arch:b}).
Finally, service providers can design their applications to offload as much business logic as possible to web clients (\Cref{fig:webapp-arch:c}), which has three main advantages:
\begin{inlinelist}
	\item lower costs for service providers, as the use of client-side resources does not generate any costs;
	\item higher responsiveness for users, as less interactions with service providers are necessary; and
	\item improved scalability, as the whole application scales with increasing client numbers.
\end{inlinelist}
This offloading is usually implemented with client-side scripting languages in the form of JavaScript and -- more recently -- WebAssembly~(see~\Cref{sec:bg:webassembly}), a binary instruction format designed for the web.

However, the aforementioned approaches, always come at certain costs~\cite{kondo2009cost,hellerstein2018serverless}, which can become a true obstacle for start-ups or non-profit organisations; but also large companies might want to reduce their cloud bills (\eg \cite{apple-aws-costs}).
Thus, outsourcing application logic to web clients seems the most promising option for service providers.
In fact, 97\% of websites do use client-side languages for offloading~\cite{share-js}.
However, this approach is limited by three fundamental obstacles.
First, web clients are not trustworthy:They can return wrong or incomplete results and can, therefore, not be used to the full extent.
This is especially relevant if the executed code or processed data contains sensitive information of the service provider or other users.
Second, clients so far perform computations only for themselves despite their potentially idle resources.
Third, more and more clients are battery-powered with less capable CPUs and can therefore not be used for extensive computations, for example wearables like watches or glasses.

To unleash the full potential of client-side computing in web applications, \magic transforms browsers into distributed points of execution.
We apply trusted execution technology to overcome untrusted clients and support cloud fallback in the case of insufficient browsers being available (\Cref{fig:webapp-arch:d}). 
In this paper, we present \textbf{\magic}, a novel web framework that allows service providers to freely distribute their web application over its currently active users.
\magic adopts the programming model of \ac{FaaS}, where applications are decomposed into self-contained functions which can be scheduled and scaled independently.
By transparently distributing these functions across participating browsers, a \magic web application can be actively provisioned by their users.
The remainder of this paper is organised around \magic's motivation and contributions:

\begin{myitemize}
\item [\S\ref{sec:bg}] introduces the main technologies behind  \magic and discusses the threat model that we consider when offloading function execution to untrusted browsers;
\item [\S\ref{sec:design}] describes \magic's design, explaining the systems requirements, the systems orchestration and how trust in remote function execution is established;
\item [\S\ref{sec:impl}] gives details on how we implemented \magic; and finally, 
\item [\S\ref{sec:eval}] evaluates \magic and shows that it scales linearly with increasing numbers of clients, copes well with unresponsive clients and performs better than costly cloud deployments.
\end{myitemize}

%

\section{Offloading Functions to Web Browsers}
\label{sec:bg}
To pave the way for trusted client-side computing in web browsers, we describe three emerging technologies that form the basis of \magic.
Since it offloads function execution to nearby web browsers, peer-to-peer communication between those is beneficial in terms of latency (\Cref{sec:bg:webrtc}).
For the computation itself, we make use of browser-based computation by using JavaScript and WebAssembly~(\Cref{sec:bg:browsercomputation}).
For establishing trust between the web browsers, we use Intel SGX~(\Cref{sec:bg:sgx}).
Finally, we conclude this section with defining our threat model~(\Cref{sec:bg:threatmodel}).

\subsection{Browser-to-browser Communication}\label{sec:bg:webrtc}
In the past, browsers were unable to communicate directly with each other.
Instead, communication was achieved by using the server-side web application as a relay; adding unnecessary latency and imposing a potential bottleneck.
\acf{WebRTC}~\cite{webrtc} changes this situation by defining a collection of protocols that enable direct browser-to-browser communication and provide transparent fallback support for restricted network settings.
\webrtc is already supported by all major browsers and provides mature JavaScript APIs.
In addition to audio and video streams, it offers \emph{data channels} to transmit arbitrary data.
%
\webrtc cannot work entirely without servers: For connection establishment, a well-known \emph{signaling server} is needed.
This server relays messages from clients for connection establishment.
\magic uses \webrtc to invoke distributed functions directly from other browsers.



\subsection{Computation in Web Browsers}
\label{sec:bg:browsercomputation}
Active content within web pages has already been introduced in 1996 with JavaScript, allowing arbitrary computations.
However, the language itself has many shortcomings, \eg its weak type system.
Therefore, JavaScript is increasingly being used as a compilation target for stricter programming languages such as TypeScript~\cite{bierman2014understanding}.
\label{sec:bg:webassembly}
The newest advancement in terms of browser-based computing is WebAssembly, a platform-independent binary instruction format~\cite{haas2017bringing, wasm_spec}.
The main goal of WebAssembly is a portable code for a safe, high-performance execution environment in web browsers.
Unlike JavaScript, WebAssembly code is not designed to be written by programmers directly; instead, it is used as a compilation target.
While the Emscripten~\cite{zakai2011emscripten} toolchain provides mature support for C/C++, wasm-pack~\cite{wasm-pack} targets Rust compilation.
Furthermore, support for other programming languages including C\#, Java, Go, and Python is currently being developed~\cite{wasm-langs}.
In \magic, the JavaScript and WebAssembly runtime V8~\cite{v8} is used as a client-side execution environment.

\mypar{Trusted Computation in Web Browsers}
\label{sec:bg:trustjs}
From a service providers point of view, computations performed in browsers are untrusted.
Browsers can return wrong, incomplete or no results at all.
This problem is usually circumvented by validating the results on the server-side, which often involves recomputations~\cite{vikram2009ripley}.
Additionally, it has been shown that validation of input values on both client and server-side can introduce vulnerabilities~\cite{bisht2010ccs,alkhalaf2014issta}.
Furthermore, this approach leads to undesired code duplication between client- and server-side code; in the worst case in different programming languages.
These duplications can be prevented by establishing trust into web browsers.
There are two research projects that achieve this by employing trusted execution technology: \trustjs~\cite{goltzsche17trustjs} and Fidelius~\cite{eskandarian2018fidelius}.
However, both neither support direct interaction between browsers nor utilise WebAssembly, essential features for \magic.

\subsection{Intel Software Guard Extensions}
\label{sec:bg:sgx}
Starting 2015, Intel's consumer CPUs include the \acf{SGX}, which enable the instantiation of \acp{TEE}.
A \ac{TEE} protects the integrity of code and data and is called an \emph{enclave}.
All computations inside such an enclave are isolated from potentially malicious software components, including privileged code such as the \ac{OS}/hypervisor or other enclaves.
Enclaves are always attached to a process and allocate an isolated memory region within its address space.
Pages of this memory region are stored in a reserved memory region, called the \ac{EPC}.
\ac{SGX} protects the integrity and confidentiality of all \ac{EPC} pages using checksums and transparent memory encryption.
Additionally, enclaves can be authenticated by a remote challenger in a process called \emph{remote attestation}~\cite{anati2013innovative}.
Thereby, an enclave identity including the enclave's code and optionally user-defined data is authenticated. 
The challenger sends the signed report to a trusted attestation service that can confirm the trustworthiness of the corresponding enclave, \ie whether it contains the expected code and is running on a genuine SGX-enabled platform.
This feature is useful, when enclaves are executed on remote infrastructure.

\mypar{Adopting Applications for SGX}
Existing applications cannot be executed in SGX enclaves without additional measures: Due to the underlying threat model, certain operations like system calls are not allowed, as they might compromise enclave isolation.
The straightforward way of enabling legacy applications (\eg a WebAssembly runtime) to run in SGX enclaves is to manually partition them (\eg\cite{brenner2016securekeeper}).
However, this imposes an extensive effort for larger applications and only limited approaches for automated partitioning exist~\cite{lind2017glamdring,tsai2020civet}.
Several research projects~\cite{arnautov2016scone,tsai2017graphene,priebe2019sgx} already explored possibilities of running legacy applications without changing their code using shielded execution and library \acp{OS}.
\magic uses such a library \ac{OS}, namely SGX-LKL~\cite{priebe2019sgx} to enable execution of JavaScript and WebAssembly inside SGX enclaves.

\mypar{Availability and Alternatives}
Being available on all recent Intel CPUs usually found in laptops and desktops and originally considered a client-side technology, our implementation of \magic naturally relies on SGX.
However, \magic's design is not tied to it and alternative TEE implementations (\eg TrustZone~\cite{armtz} or Keystone~\cite{lee2020keystone}) could be used.
In \Cref{sec:other-tee}, we discuss how \magic can be implemented on other \ac{TEE} platforms.

\subsection{Threat Model} \label{sec:bg:threatmodel}
In a \magic deployment, two entity types interact frequently: many \emph{users} and a single \emph{service provider}.
The users want to use the web application provided by the service provider.
In turn, the service provider wants to make use of the users spare resources to provide that very service.
Since client machines are out of control of the service provider, they are considered as untrustworthy and treated as potential attackers.
We propose that providers offload sensitive computations or critical data only to clients equipped with a \ac{TEE}. 
Furthermore, users do not trust the machines of other users.
However, we assume the user to trust the service provider to
\begin{inlinelist}
	\item provide a benign service; and
	\item perform proper remote attestation (see \Cref{sec:bg:sgx}) with all other connected peers.
\end{inlinelist}
The code that clients receive is regarded as untrusted, an assumption which is equivalent to current best practices in the web:
Although users trust service providers, \eg when sending private data, delivered code is still executed in a sandbox.
Also, the remote attestation assumption is reasonable, as a proper attestation is in the interest of the service provider.
Finally, the resources provided by the users are still unreliable:
Frequent disconnects have to be expected, as users might close their browsers, shut down their machines or have an unstable network connection.
Since \magic relies on Intel SGX, we assume a correct implementation of SGX in hard- and software as well as a properly working attestation service (see \Cref{sec:bg:sgx}).
We are aware of side-channel attacks affecting SGX~\cite{weichbrodt2016asyncshock,lee2017inferring,van2018foreshadow,kocher2018spectre,lsds2018spectresgx,chen2019sgxpectre,ZombieLoad2019,van2020lvi,murdock2020plundervolt}.
These have either been fixed by updates of microcode~\cite{intel2018l1tf} or the SGX SDK~\cite{intel2018spectreadvisory} or mitigated~\cite{shih2017t,oleksenko2018varys,DBLP:journals/corr/abs-1712-08519,intel_mitigation_whitepaper}.
We therefore acknowledge potential side-channel attacks as a concern, but ultimately consider them out of scope of our work.
In contrast to most SGX-related threat models (\eg in \cite{Schuster:2015:vc3,brenner2016securekeeper,arnautov2016scone,weiser2017sgxio,priebe2018enclavedb,goltzsche2018endbox}), we explicitly do consider denial-of-service attacks in the distributed setting of \magic: If parts of the application are subject to such attacks, execution can be transferred to other participants or the server-side fallback environment.

\newcommand{\reqIntPriv}{$R_1$\xspace}
\newcommand{\reqLangAgno}{$R_2$\xspace}
\newcommand{\reqIsolation}{$R_3$\xspace}
\newcommand{\reqReliability}{$R_4$\xspace}
\newcommand{\reqScalability}{$R_5$\xspace}
\newcommand{\reqPerformance}{$R_6$\xspace}

\section{Design}
\label{sec:design}
In this section, we describe how we design \magic to transform web browsers into distributed points of execution for web applications.
We start by detailing the system requirements in accordance with the threat model defined above in \Cref{sec:design:req}, followed by a brief overview of \magic in~\Cref{sec:design:nutshell}. 
Furthermore, we give details on how \magic is deployed on server- (\Cref{sec:design:server-side}) and client-side (\Cref{sec:design:client-side}), discuss which types of applications benefit most from \magic (\Cref{sec:design:applications}) and describe security-related design choices (\Cref{sec:design:security}).

\subsection{System Requirements}
\label{sec:design:req}
A system for securely offloading functions into distributed, untrusted web browsers must fulfil the following requirements:

\begin{enumerate}[leftmargin=14pt, itemindent=0pt, labelwidth=0pt]
 \item[\reqIntPriv] \emph{Function integrity and confidentiality.} \magic should protect the integrity of all functions as well as the confidentiality of their in- and outputs.
 \item[\reqLangAgno] \emph{Language agnosticism.} \magic should support a wide range of programming languages for developers to choose from, also to ease the porting effort for legacy applications.
 \item[\reqIsolation] \emph{Isolation.} Function execution must be isolated from the executing host, \ie, the code delivered to clients needs to be unable to access or even manipulate their environment.
 \item[\reqReliability] \emph{Availability.} All function calls issued to \magic should eventually succeed, even if insufficient browsers participate or individual browsers misbehave.
 \item[\reqScalability] \emph{Scalability.} \magic should scale well with an increasing and fluctuating number of clients.
\end{enumerate}

\noindent In the following, we describe \magic in a nutshell to explain how it fulfils these requirements.

\subsection{\magic in a Nutshell}\label{sec:design:nutshell}
\begin{figure}
	\includegraphics[width=0.85\columnwidth]{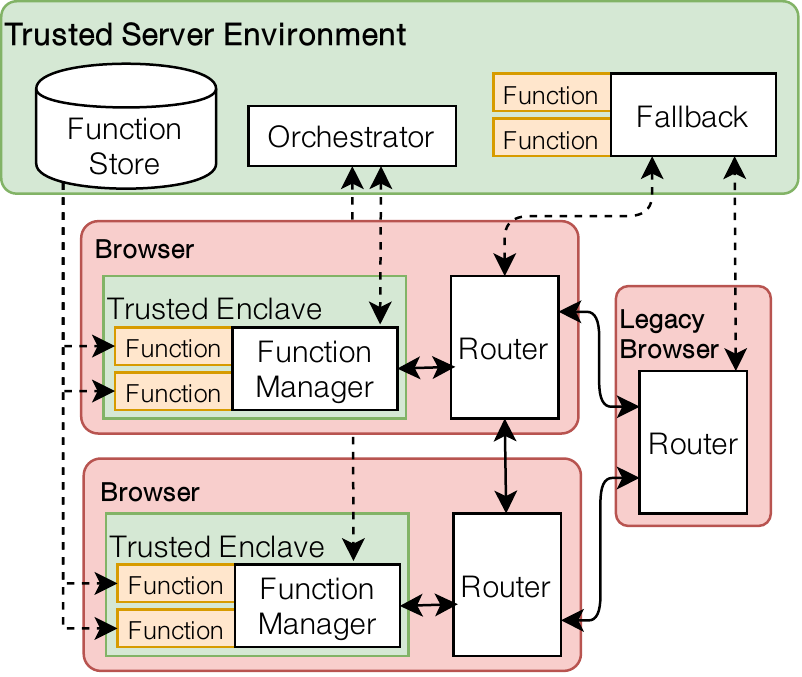}
	\caption{Simplified architecture of \magic with two \magic browsers and one legacy browser. Trusted components are denoted in green, untrusted ones in red. Solid arrows show interactions for function invocations, while dashed arrows show more infrequent interactions such as fetching functions, communication with the orchestrator or fallback.}
	\label{fig:simple-arch}
\end{figure}
\Cref{fig:simple-arch} shows the simplified architecture of \magic with three clients; two clients with the \magic browser (\ie a browser with support for trusted function execution) and one legacy browser.
\magic allows deploying web applications on distributed, usually untrusted browsers to reduce the number of centralised servers needed for operation.
All \magic browsers that load a \magic application automatically contribute to provisioning that very application, thus, increasing overall scalability~(\reqScalability).
Legacy browsers (\eg clients without \ac{TEE} support) can still benefit from other \magic browsers, but cannot contribute to the application.
Following the programming model of \ac{FaaS}, a \magic application is decomposed into \emph{functions}, which allows a fine-grained, parallel distribution, also increasing scalability~(\reqScalability).
\magic functions can be developed using JavaScript and WebAssembly~(see \Cref{sec:bg:browsercomputation}), which provide two important properties:
First, isolation from the host~(\reqIsolation) is guaranteed, because runtimes for both languages are designed as execution sandboxes, which are already used for executing untrusted code in browsers.
Second, a wide range of programming languages is supported~(\reqLangAgno), as both languages can act as compilation targets (see \Cref{sec:bg:browsercomputation}).

When a \magic browser loads a \magic application, the framework first contacts the \emph{orchestrator}.
The orchestrator has a global view over all participating browsers and can decide which set of functions the newly connected browser should \emph{install} locally, based on an application-specific policy (\eg highest priority for latency-sensitive functions).
For installation, function code is first fetched from a \emph{function store} and then deployed locally in a \ac{TEE} based on Intel SGX~(see \Cref{sec:bg:sgx}).
This \emph{trusted enclave} protects the integrity of function code and confidentiality of in- and outputs (\reqIntPriv).
Optionally, functions can be delivered in an encrypted format to gain code confidentiality.
%
Installed functions can then be \emph{invoked} by other browsers.
The \emph{router} located on every client forwards the application's \emph{function invocation request} to clients that have that function installed.
Note, that this can also be the client that issued the request.
All functions are additionally deployed in the server-side \emph{fallback} component, which can be based on a traditional \ac{FaaS} infrastructure.
This allows bootstrapping a \magic application without any peers being available~(\reqReliability).

Providers of a \magic application should try to distribute the workload evenly across clients, so that users cannot differentiate between provisioning and using an application.
Depending on the type of application, altruistic users might directly donate resources, such as in volunteer computing systems.
For other applications, premium models can be used to create incentives, \eg by offering contributors ad-free services, exclusive features, or digital goods.
Existing approaches~\cite{tople2018vericount,alder2018s,goltzsche2019acctee} can be used to perform resource accounting on client-side and attest service providers that a certain amount of computation has been performed.


\subsection{\magic on Server-side}\label{sec:design:server-side}

Based on \Cref{fig:simple-arch} we here describe \magic's server-side components and discuss deployment strategies for these.

\mypar{Orchestrator}
\label{sec:design:orchestrator}
The orchestrator keeps track of all available peers and can instruct them to load a set of functions.
Additionally, newly connecting clients can request information about available peers.
Before deployment, the service provider adds metadata to each function: a \emph{distribution factor} and a \emph{weight}.
The higher the distribution factor, the more will the corresponding function be installed across available peers.
The weight represents the \emph{expensiveness} of functions, reflecting how many resources it uses. 
When \magic is initialised on a new client, it connects to the orchestrator to announce its availability and \emph{machine size}, reflecting its computing power.
This value can be determined by different approaches:
\begin{inlinelist}
	\item the browser's user agent,
	\item static hardware information, and
	\item a benchmark executed before connecting.
\end{inlinelist}
Subsequently, the orchestrator identifies the \emph{least represented function} fitting the connecting client according to distribution factor and weight and instructs it to install this function.
This process is repeated, as long as the cumulative weights of all functions does not exceed the machine size.
The orchestrator persists the information which peer has which function(s) installed to forward it to new clients.
When a \magic application needs to invoke a function, the \magic client first sends a request to the orchestrator containing the unique function name.
The orchestrator responds with a list of peers that have already loaded this function.
This list is limited by proximity of peers, which can be determined by geotargeting or estimated latency to the orchestrator.
It is cached for 60 seconds by the \magic client to prevent it from contacting the orchestrator too often.

\mypar{Function Store}\label{sec:design:functionstore}
The function store serves static code in form of JavaScript or WebAssembly that makes up the application using \magic.
In order to assert trust, all code is integrity protected and can optionally be encrypted.

\mypar{Fallback} \label{fallback}
For every \magic application, it is essential that sufficient peers are available.
However, especially for newly released applications, this might not always be the case.
\magic includes a fallback that is installed on trusted infrastructure.
We expect the resource demand of the fallback to be low, as only very few users would need to make use of it, therefore, deploying it as \ac{FaaS} functions can reduce costs.

\mypar{Deployment Strategies}
Consisting of multiple components and involving different parties, \magic applications can be deployed in different ways.
In the following, we describe strategies that can improve scalability and latencies. 
%
The client-side code is delivered to the browsers contributing to the \magic application via HTTP. 
As trusted function code is static, and integrity protected, serving can be delegated to a \ac{CDN} to improve scalability and end-user latency.
Furthermore, all \magic clients interacting with the orchestrator makes it a potential bottleneck.
However, as clients are only interested in peers within close proximity, multiple orchestrators can be deployed in different regions where every orchestrator is responsible for a disjoint set of peers.
This approach is transparent to clients and reduces the load of and the latency to orchestrator nodes. 

\newcommand{\pA}{$\lgwhtcircle\lgwhtcircle\lgwhtcircle$}
\newcommand{\pB}{$\lgblkcircle\lgwhtcircle\lgwhtcircle$}
\newcommand{\pC}{$\lgblkcircle\lgblkcircle\lgwhtcircle$}
\newcommand{\pD}{$\lgblkcircle\lgblkcircle\lgblkcircle$}
\begin{table*}
\begin{tabular}{lclllc}
\toprule
application category & example app & data source & data sink & data size & suitability\\
\midrule
collaboration tool & \cite{etherpad} & peer & peer & small & \pD\\
web game & \cite{rooney2004federated} & peer & peer & small & \pD\\
\textbf{digital assistant} & \cite{campagna2017almond} & peer & cloud & small & \pD\\
private web search & \cite{pires2018cyclosa} & peer & cloud & small & \pD\\
upload preprocessing & \cite{twitter}  & peer & cloud & medium & \pC\\
\textbf{recommendation system} & \cite{davidson2010youtube} & cloud & peer & small & \pC\\
video conferencing & \cite{ivov2013hangout} & peer & peer & large & \pC\\
p2p video streaming & \cite{da2019privatube} & peer & peer & large & \pC\\
video transcoding &  & cloud & cloud & large & \pB\\
\bottomrule
\end{tabular}
\caption{Suitability of different web application categories for \magic. Categories showcased in this paper are highlighted.} \label{tab:apps}
\end{table*}

\subsection{\magic on Client-side}\label{sec:design:client-side}
\begin{figure}
	\includegraphics[width=0.8\columnwidth]{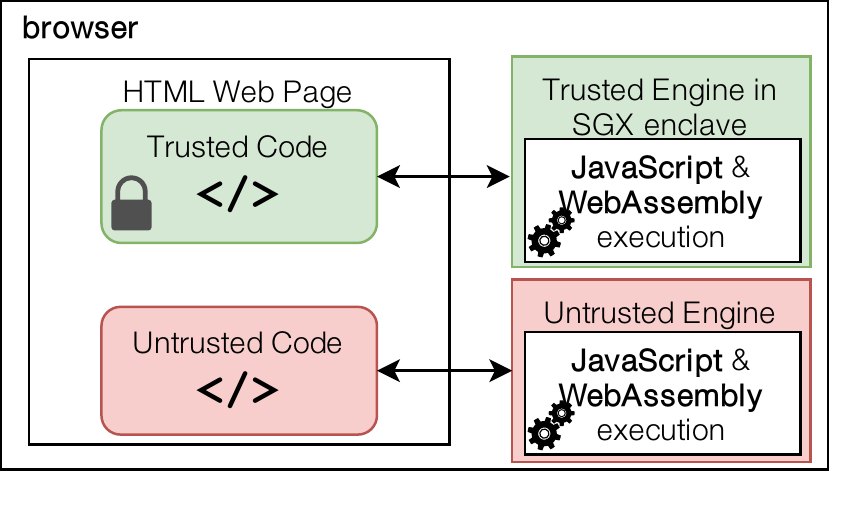}
	\caption{Architecture of the \magic browser with trusted execution capabilities.}
	\label{trustjs-arch}
\end{figure}
As shown in \Cref{trustjs-arch}, the \emph{\magic browser} contains an additional execution engine protected by an SGX enclave, therefore being trusted.
Trusted JavaScript or WebAssembly code is marked using special \acs{HTML} tags for identification.
Additionally, the code is signed and optionally encrypted before being delivered to peers.
Before execution, the code is moved into the trusted engine, where it gets verified and decrypted.
The \magic browser also generates code for proxies to transparently handle calls to and from the trusted execution environment.
The client-side components of \magic are separated into a trusted and untrusted part.
Functions are executed in the trusted part, which can happen on behalf of a local or remote function invocation.
In the untrusted part, the router uses information from the orchestrator to delegate function invocations.

\mypar{Starting \magic}
As described earlier, the orchestrator instructs newly connected clients to load a specific set of functions after loading a website.
The client's untrusted side downloads these functions from the function store and calls into the trusted part to forward the  code.
In the trusted part of the application, the \emph{function manager} verifies the signature and decrypts the code if necessary; then it installs the functions.

\mypar{Connection Management} \label{conn-man}
After requesting peers from the the orchestrator, the router preemptively establishes \ac{WebRTC} connections to all of them to decrease latencies for first-time calls.
For instance, in a web application where the user needs to log in before doing any actions, the connection to a peer with the login function can already be established before users finish entering their credentials.
The first invocation is issued to the first established connection to reduce call latencies.
After this invocation, clients measure the latency to choose low-latency peers for future calls. 
During application runtime, peers are expected to suddenly disappear, \eg due to network issues or users closing their browsers.
Therefore, peer connections are terminated upon close requests or if \webrtc keep-alive messages are not responded to.
When many peers disconnect, the \magic client asynchronously issues new requests to the orchestrator.

It is possible that the client sends a request to a peer before noticing that that peer has become unavailable.
Whenever a remote function invocation times out, the router terminates the connection to that peer and uses the next best peer for function invocation.
The timeout duration is configurable per function and is dependent on the workload, \eg long running functions need a higher timeout.


\subsection{\magic Functions}
The parts of \magic applications that should be offloaded to nearby browsers need to be self-contained functions.
Applications can instruct the function manager to invoke a function by passing its name and parameters.
The function manager directly invokes it, if it is installed locally, otherwise, a remote peer is contacted.
For that, the router uses already established functions to peers that have the function installed and sends an \emph{invocation request} via \ac{WebRTC}, including the encrypted parameters.
After the receiving peer's function manager decrypts the parameters, the function is invoked locally and the encrypted result is returned to the client.
In the case of no peers with the appropriate function being available, the fallback is used for function invocation.
For the application, function invocation is transparent, regardless of where the function is ultimately invoked.
Additionally, functions can also call other functions, no matter if they are installed locally or remotely on other peers.

The router manages a \emph{peer rating} for every peer by periodically measuring the latency to all connected peers.
Using the peer with the lowest latency allows \magic to find the closest peer available.
Additionally, overloaded peers are automatically relieved when their latency rises.
The decision which peer to use is repeated on every remote function invocation.

\mypar{Function State}
\magic functions are -- in the best case -- stateless, \eg they do not rely on persistent data shared with other functions.
However, function code can access existing storage services for sharing data.
This approach represents the standard in the cloud computing industry (\eg AWS Lambda functions accessing S3 storage).
\magic also supports direct access to services that are usually part of the back-end (\eg a database, blob storage or key-value store) as described in \Cref{sec:design:proxy}.
Additionally, \magic supports \emph{local function state} by allowing functions to persist data in browsers.



\subsection{\magic Applications} \label{sec:design:applications}
Certain types of web applications are more suited for a \magic deployment than others.
For example, \magic is especially well-suited for applications that do not depend on centrally stored data as input.
Additionally, the size of processed data is important, also whether it is consumed on \magic peers or stored centrally.
In general, \magic improves latency for end-users and helps service providers to reduce costs by offloading computations.
Therefore, latency-sensitivity and computation-intensiveness influence how suitable a given application is for \magic.
\Cref{tab:apps} lists several example application categories, showing their main data sources (excluding initialisation) and data sinks.
Here, \emph{peer} means that data is produced or consumed on \magic peers, whereas \emph{cloud} means that data is loaded from or stored at central locations.
Furthermore, the table distinguishes between data sizes from small ($<$10\unit{MB}) over medium ($<$500\unit{MB}) to large ($\geq$500\unit{MB}).
We rate the applications suitability for \magic with excellent~(\pD), good~(\pC) or poor~(\pB).
For example, applications that consume small amounts of data from peers are excellent use cases for \magic, such as collaboration tools , web games, digital assistants or private web search.
We still assume good suitability, if applications process medium data sizes generated at peers, \eg for preprocessing uploads such as images or videos for social networks.
The same applies for applications where small amounts of cloud-based data is processed, such as in a recommendation system.
When processing large amounts of data, we rate the application to still have good suitability, if the data is produced and consumed on \magic peers, as in peer-to-peer video conferencing streaming.
However, when a lot of cloud-based data is processed and then again stored in the cloud, we see poor suitability, \eg video transcoding for cloud-based streaming services.

\subsection{Security} \label{sec:design:security}
Service providers of \magic applications need to ensure that only \ac{SGX}-enabled clients execute functions, and, that these clients execute the correct code.
To achieve that, the remote attestation (see \Cref{sec:bg:sgx}) facilities of SGX are used.
While loading the web page, a remote attestation request is sent to the service provider, containing two important assets:
\begin{inlinelist}
	\item a \emph{quote}, which identifies the code of the SGX enclave and can be forwarded to the \ac{IAS}; and
	\item an ephemeral public key that is generated during enclave startup.
\end{inlinelist}
Upon retrieval, the service provider verifies that the enclave measurement matches a pre-known value, thus indicating that the client is running a genuine \magic enclave containing the expected code.
Then, the quote is forwarded to the \ac{IAS}, a positive reply indicates that the enclave was started on a genuine SGX-capable platform.
The remote attestation process finishes with the service provider encrypting a symmetric cryptographic key called the \emph{script secret} with the enclave's public key.
This key can be updated regularly and is unique for every web application.
\magic clients use this secret to verify \acp{MAC} and to decrypt the trusted JavaScript code embedded in the \acs{HTML} of the application and 
JavaScript code or WebAssembly modules retrieved from the function store.

\mypar{Attestation between peers}\label{sec:design:attestation}
In \magic, enclaves mainly communicate with other enclaves, \eg for function invocation.
All communication is integrity protected and encrypted using the script secret.
The service provider is considered trusted (see \Cref{sec:bg:threatmodel}) and only exposes the script secret to trustworthy enclaves.
Therefore, \magic establishes trust between peers using the script secret because
\begin{inlinelist}
	\item only trustworthy enclaves are able to decrypt peer-to-peer messages, and
	\item an enclave receiving such a message can be sure that the communication partner has undergone a successful remote attestation with the service provider.
\end{inlinelist}
Using this \emph{implicit attestation}, no further actions after connecting to a peer
need to be performed to establish a secure channel.

\mypar{Generic TLS Proxy}\label{sec:design:proxy}
Most web applications interact with back-end systems for server-side application logic or storage (\eg blob storage, databases or key-value-stores).
These systems are usually deployed behind web servers, should not interact with clients directly, and often have no support for encrypted access.
Therefore, \magic includes the \emph{generic TLS proxy}, which is deployed in front of back-end systems and uses a special \ac{CA} to enable \emph{implicit remote attestation} between the proxy and connected enclaves.
The \ac{CA} issues client certificates after the remote attestation process described before.
Accepting only certificates from this \ac{CA}, the proxy employs mutual authentication to ensure only genuine \magic enclaves can access the back-end systems.



\section{Implementation}\label{sec:impl}
In this section, we give details on how \magic was implemented on server- and client-side.

\mypar{Server-side}
The function store is implemented as an HTTP server in TypeScript, the orchestrator is written in TypeScript as well.
The signaling server uses \emph{PeerJS}~\cite{peerjs} for \webrtc signaling based on the WebSocket protocol.
We implement the orchestrator and signaling server in the same process; as this eases communication between these components, \eg for notifications when clients disconnect.
The fallback (see~\Cref{fallback}), can be deployed using Node.js as the server-side runtime or via existing FaaS architectures; we use \emph{Cloudflare Workers}~\cite{cloudflareworkers}, as they already support WebAssembly~\cite{cloudflareworkerswasm} and \emph{AWS Lambda}~\cite{aws-lambda}.
\magic's generic proxy~(see~\Cref{sec:design:proxy}) is based on the universal TLS tunnel \emph{stunnel}~\cite{stunnel} and is implemented as an HTTP server in Node.js.

\mypar{Client-side}
The client-side part of \magic is written in TypeScript as well, connects to the orchestrator via WebSockets and uses PeerJS for \webrtc support.
%
Although we envision \magic as a web standard that is natively supported by browsers, we resort to implementing it as a browser extension based on the \emph{WebExtensions} API~\cite{webextensions-api}.
The extension communicates via native messaging~\cite{native-messaging} with the \emph{backend}, the V8~\cite{v8} runtime on top of SGX-LKL~\cite{priebe2019sgx} for JavaScript and WebAssembly execution.

\mypar{Applicability on Other \ac{TEE} Platforms} \label{sec:other-tee}
The only SGX-specific component of \magic is this backend.
Replacing it with an implementation based on another technology is possible and porting \magic to other \ac{TEE} platforms should be effortless.
Besides SGX, ARM TrustZone~\cite{armtz} is the most spread TEE implementation, especially on mobile devices.
For example, OP-TEE~\cite{optee} or Knox~\cite{knox} could be used for porting \magic.
In this case, not only the hardware mechanisms must be trusted, but also the hardware vendor, \eg Samsung.
However, every service provider could individually decide which TEE implementation to trust.



\section{Evaluation}
\label{sec:eval}

In this section, we evaluate different aspects of \magic and its underlying technologies.
We start with evaluating \magic's latency impact (\Cref{sec:eval:sgx-lat}).
Then, we describe use case applications in \Cref{sec:use-case}: an assistant chat bot and a movie recommendation system.
We  continue with evaluating end-user latency and service costs (\Cref{sec:eval:lat-costs}) as well as the scalability of \magic (\Cref{sec:eval:invocation}).
Our last evaluation (\Cref{sec:eval:reconn}) covers the impact on unresponsive peers on \magic.
Finally, we conclude this evaluation section by discussing attacks in \Cref{sec:attacks} .

\subsection{Latency Impact of \magic}\label{sec:eval:sgx-lat}
Calls into the trusted execution environment of \magic are subject to delays from different sources such as
\begin{inlinelist}
	\item overhead of entering SGX enclaves;
	\item overhead due to copying data to/from enclaves;
	\item additional overhead induced by SGX-LKL; and
	\item overhead added by the \magic browser extension.
\end{inlinelist}
To evaluate the overall delay of \magic, we use two machines equipped with an SGX-capable Intel Core i7-6500U CPU connected via a $1\unit{GBit}$ switch.
On both machines we run the Chromium browser with our extension installed and exchange messages via \webrtc synchronously, \ie the next message is sent when the reply for the first one is received.
We send 100 messages for payload sizes from $16\unit{bytes}$ to $1\unit{MB}$.
To compare \webrtc via \magic with plain \webrtc, requests and replies are either routed through the \magic browser extension and the attached SGX enclave or not.
As baseline, we use ICMP pings with sizes from $16\unit{bytes}$ to $32\unit{KB}$, as larger payloads are not possible.
Furthermore, we add artificial network delays of $10\unit{ms}$ and $50\unit{ms}$ using \texttt{tc netem} to simulate different distances.
\begin{figure}
\includegraphics[width=\linewidth,page=1]{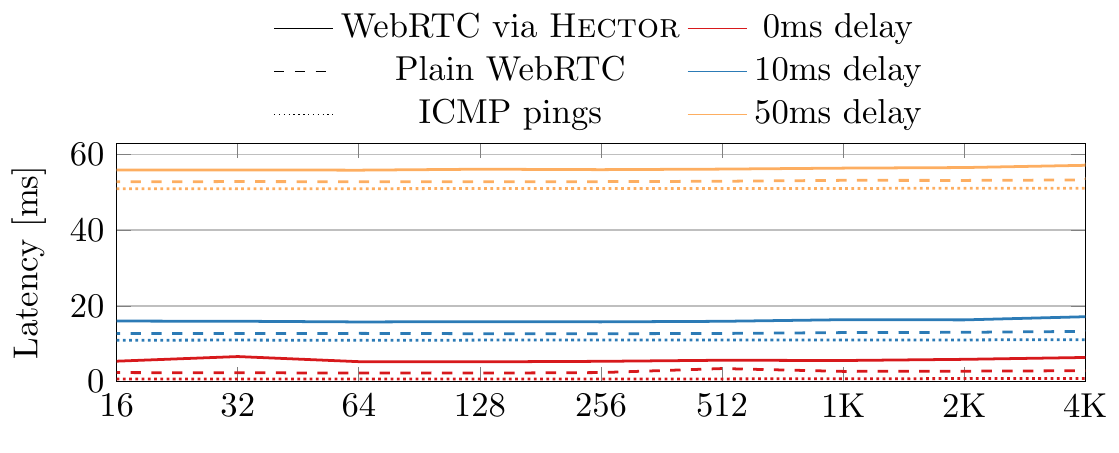}
\includegraphics[width=\linewidth,page=2]{plots/sgx-lat.pdf}
\caption{Average latency impact of \magic for different network delays.}
\label{fig:sgx-lat}
\end{figure}
\Cref{fig:sgx-lat} shows the average latencies for one message exchange.
We ignore the reproducible outlier for \webrtc via \magic with a $10\unit{ms}$ delay for a $256\unit{KB}$ payload, as we suspect a bug in either Chromium or \magic's browser extension.
For smaller payloads up to $4\unit{KB}$, the latencies are constant, with plain \webrtc adding $1-2\unit{ms}$ and \magic additionally adding  $3-4\unit{ms}$.
For larger payloads up to $1\unit{MB}$, the latencies increase; \magic adds up to  $250\unit{ms}$ due to larger buffers needing to be copied from the website via the browser extension into the SGX enclave and back.
For no artificial delay (\ie for sub millisecond latencies), the relative latency overhead of \magic is high: $2.2\times$ for small and $68\%$ for large payloads.
For medium artificial delays ($10\unit{ms}$), \magic's average latency overhead for all payload sizes is $22\%$.
Finally, for larger artificial delays ($50\unit{ms}$), the average latency overhead is $6\%$ for small payloads and up to $20\%$ for larger payloads.
Note, that compared to average page load times of popular websites in the order of hundreds of milliseconds up to several seconds~\cite{wang2013demystifying,goltzsche2018endbox}, this overhead is negligible. 

To summarise, \magic adds a small delay in most cases.
While this is a relatively large overhead in a local network, it does have a small effect on web- or cloud-based use cases with larger latencies.
Note, that the browser extension itself induces the highest overhead. 
If trusted execution would be natively supported by browsers, the only overhead would be the latency penalty of enclave calls in the order of microseconds~\cite{weisse2017regaining,weichbrodt2018sgx}.

\subsection{Use Cases}\label{sec:use-case}

To evaluate \magic we implement two realistic web-based use cases with different suitabilities from our discussion in \Cref{sec:design:applications}:
First, a digital assistant based on machine learning which classifies text inputs and, second, a movie recommendation system.
We intentionally choose these two quite different use cases:
While the digital assistant processes data produced at the peers and does not process centralised data at all, the recommendation system depends on a centralised database.
However, both process relatively small amounts of data.

\mypar{Digital Assistant}
Our digital assistant is a chat bot designed to be embedded into websites to interact with its users.
It applies \ac{NLU} on text messages with a pre-trained model of approximately 5\unit{MB} to identify intents.
Seven intents are supported: \texttt{greet}, \texttt{bye}, \texttt{affirmative}, \texttt{negative}, \texttt{wtf}, \texttt{playMusic}, and finally \texttt{addEventToCalendar}.
The first four intents are used for controlling the conversational flow, while insults or out-of-context messages are classified as \texttt{wtf}.
The two last intents trigger actual actions; either streaming songs or creating calendar entries.
Our bot is implemented in TypeScript and is based on the Aida chat bot~\cite{aida}.
Under the hood, it uses Tensorflow.js~\cite{tensorflow-js}, which also supports training and classification in WebAssembly.
However, in our implementation we perform the classification in pure JavaScript (using Tensorflow's \texttt{cpu} backend), since the WebAssembly implementation is still under development and did not work with our model.
The intent classification function can be deployed on \magic peers as well as on the AWS Lambda \ac{FaaS} platform~\cite{aws-lambda}.
When invoked, the function fetches model data from S3, loads it into Tensorflow, then performs intent classification and finally returns the detected intent together with a confidence value between 0 and 1.
Fetching of model data only happens during the first invocation, as the model data is cached for future invocations.

\mypar{Movie Recommendation System}
The movie recommendation system allows its users to browse or rate movies.
Additionally, they can list recommended movies, while recommendations are based on their and other users' ratings.
We implement it as a \ac{SPA} using TypeScript and \emph{Vue.js}~\cite{vuejs}.
%
It contains nine functions, which allow users to log in to the application, browse movie titles by genre, view and modify movie ratings, and retrieve recommendations or information about specific titles.
All functions obtain information from a central MySQL database, which is securely accessible through \magic's TLS Proxy (see \Cref{sec:design:proxy}).

\begin{figure*}[t]
\begin{minipage}{0.56\textwidth}
\begin{tabular}{lp{1.7cm}p{1.7cm}p{1.7cm}}
\toprule
function deployment & end-user latency & most billed durations & cost for $10^7$ invocations\\
\midrule
\setA: \magic same network & 196 - 285\unit{ms} & none & \$0\\
\setB: \magic same city & 246 - 329\unit{ms} & none & \$0\\
\setC: \magic same region & 249 - 330\unit{ms} & none & \$0\\
\midrule
\setD: AWS Lambda 3008\unit{MB} & 370\unit{ms} & 300 - 400\unit{ms} & \$150 - \$200\\
\setE: AWS Lambda 2048\unit{MB} & 378\unit{ms} & 300 - 400\unit{ms} & \$100 - \$133\\
\setF: AWS Lambda 1024\unit{MB} & 694\unit{ms} & 600 - 700\unit{ms} & \$150 - \$175\\
\bottomrule
\end{tabular}
\captionof{table}{End-user latencies, most billed durations in AWS Lambda and estimated invocation costs for chat bot use case.}
\label{tab:aws-costs}
\end{minipage}
\hfill
\begin{minipage}{0.4\textwidth}
\includegraphics[width=\textwidth]{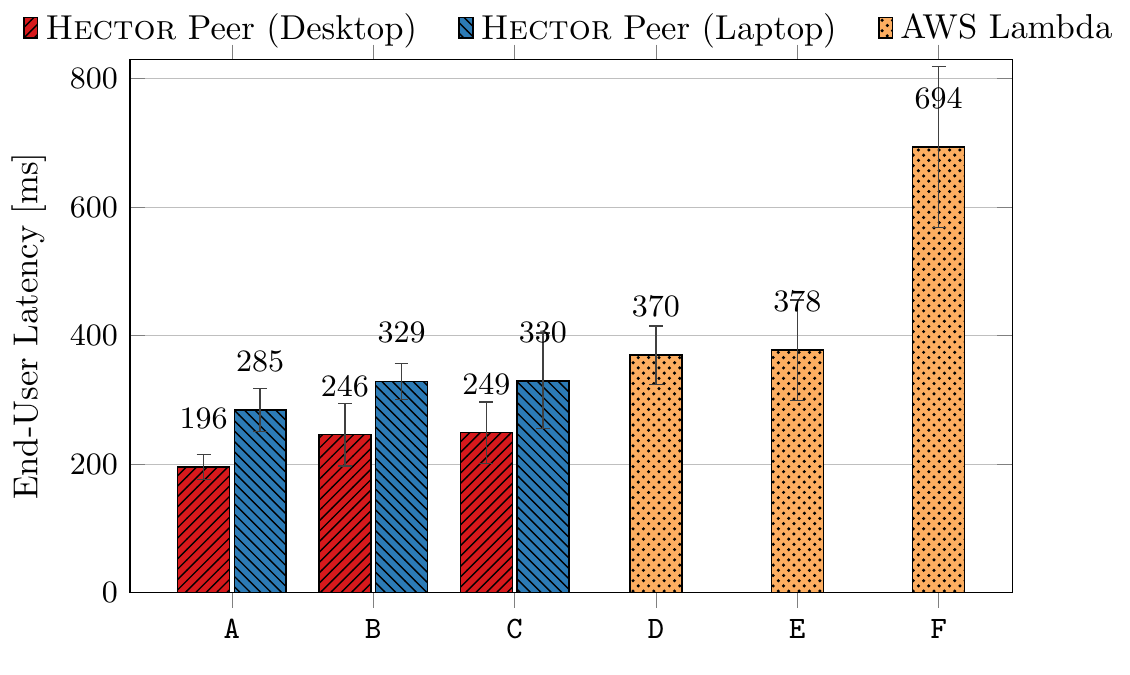}
\captionof{figure}{Average end-user latencies for the digital assistant use case for different function deployments.}
\label{fig:invoc-lat}
\end{minipage}
\end{figure*}

\subsection{End-User Latencies and Service Costs} \label{sec:eval:lat-costs}
Here, we want to show that deploying functions with \magic can actually reduce
\begin{inlinelist}
	\item the latencies experienced by end-users, and
	\item the costs for service providers.
\end{inlinelist}
For that, we compare deployments of the classification function of our digital assistant use case on \magic and AWS Lambda.
To facilitate different machine types acting as \magic peers, we install the classification function either on a laptop with an i7-6500U CPU or on a desktop machine with an i7-6700 CPU.

We deploy one invoking and one executing \magic peer in the following configurations: 
\begin{inlinelist}
	\item two \magic peers in the same university network (config \setA);
	\item two \magic peers in the same city, the invoking peer in a university network, the executing peer in a home network (config \setB);
	\item two \magic peers in the same region, the invoking peer in a university network, the executing peer in a home network in a different city in the same region (config \setC);
\end{inlinelist}
The home networks are connected to the internet via DSL.
In all configurations, the orchestrator is located in the university network, but not part of the latency measurements.
We use Node.js processes instead of browsers to perform the measurements; this allows us to run \magic in the same runtime (V8) as in most browsers while simplifying the setup.
This does only minimally influence the results, as the latency added by the \magic browser extension and its SGX enclaves is negligible (see~\Cref{sec:eval:sgx-lat}).

As a baseline, we deploy the classification function on AWS Lambda in the closest region to our university, which is \censor{\emph{eu-central-1} (Frankfurt)}.
In Lambda, developers define the allocated memory per function deployment from 128\unit{MB} to 3008\unit{MB} in 64\unit{MB} steps, which also linearly increases vCPU credits available to the function.
We deploy the function in the following configurations:
3008~\unit{MB} allocated memory, which is the maximum (config \setD);
2048~\unit{MB} allocated memory (config \setE); and
1024~\unit{MB} allocated memory (config \setF).
Lower memory amounts are not considered, because the execution durations become unacceptable (from 1 second at 512\unit{MB} up to 10 seconds at 128\unit{MB}), which additionally generates higher costs due to these extremely long durations. This is due to our function consuming approximately  500\unit{MB} of memory.

In all configurations, we use a machine in our university network for invoking the function and reporting the latency until the result is returned.
We invoke the chat bot's classification function 100 times with a randomly chosen sentences from a test data set~\cite{aida-test-data} consisting of 1500 sentences with lengths between 3 and 141 characters.
We report the average latencies experienced by end-users in \Cref{tab:aws-costs} and in \Cref{fig:invoc-lat}.
For the \magic deployments, we measure latencies between 196 and 249\unit{ms} for desktop peers and between 285 and 330\unit{ms} for laptop peers depending on placement of peers.
For AWS Lambda, we observe latencies between 370\unit{ms} and 694\unit{ms} depending on the configured Memory for the FaaS environment.
This results in latencies for \magic ranging from
$11\%$ (\setC\,laptop/\setD),
over $13\%$ (\setB\,laptop/\setE)
and $35\%$ (\setB\,desktop/\setE)
up to $410\%$ (\setA\,desktop/\setF) shorter than for AWS Lambda.

Furthermore, we estimate costs for the AWS Lambda deployments.
Function invocations are billed \$2 per $10^7$ request plus a fixed amount per 100\unit{ms} duration, rounded up to the nearest 100\unit{ms}. This fixed amount varies, depending on the AWS region and the allocated memory between 128\unit{MB} and 3008\unit{MB}~\cite{aws-lambda-prices}.
\Cref{tab:aws-costs} shows the most billed durations by Lambda per configuration and the resulting estimated costs for $10^7$ invocations for different deployments.
While function invocation on \magic peers does not generate any costs for service providers, an AWS Lambda deployment generates cost between \$100 and \$200 per $10^7$ requests.
For a web application with $10^6$ daily active users, each interacting with the chat bot 10 times, this would lead to yearly costs between \$36,500 and \$73,000.
These measurements show that function executions have significantly lower latencies compared to a traditional FaaS deployment, while generating no costs for service providers.

\subsection{Scalability} \label{sec:eval:invocation}

\begin{figure*}[h]
\begin{subfigure}{\columnwidth}
\includegraphics[width=\linewidth]{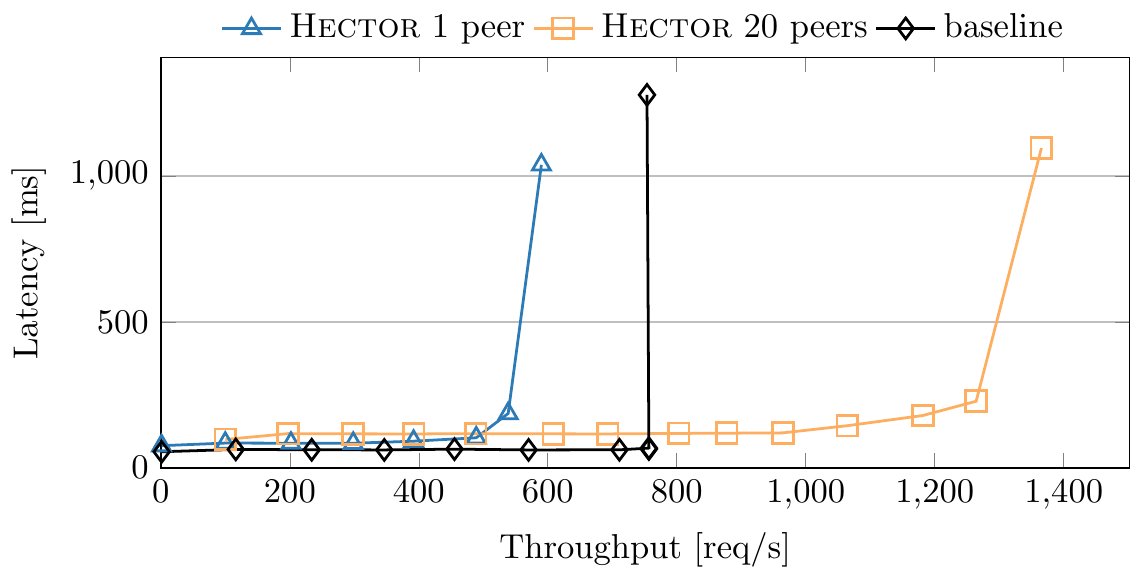}
\label{fig:tput-lat-db}
\end{subfigure}
\hfill
\begin{subfigure}{\columnwidth}
\includegraphics[width=\linewidth]{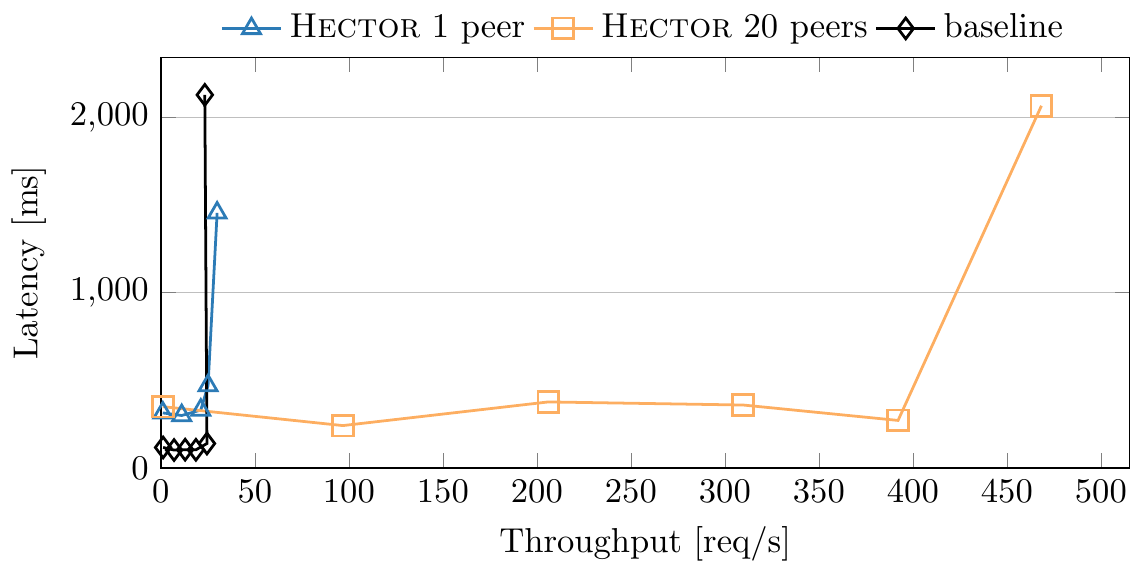}
\label{fig:tput-lat-cpu}
\end{subfigure}
\caption{Throughput and latency of invocations of I/O intensive function (left) and CPU intensive function (right).}
\label{fig:tput-lat-both}
\end{figure*}
One of the key advantages of \magic is increased scalability of web applications by leveraging client-side resources, which we want to evaluate here.
For that, we take two different functions from our movie recommendation system (see~\Cref{sec:use-case}).
First, the function \texttt{getAllGenres}, which issues queries to the database and returns a list of all genres in the database.
It is used as an example for an I/O intensive function not performing significant computation.
Second, the \texttt{getRecommendations} function, which executes a nearest neighbour algorithm for the user's ratings and 100 other users' ratings to find recommendable movies.
Although it also fetches user ratings from the database, most of its computation is used for the algorithm; it is, therefore, an example for a computation intensive function.
%
We provision different \ac{AWS} instance types (\emph{t3.micro} for peers and \emph{t3.medium} for servers, see~\cite{aws-instance-types}) and deploy one server instance for \magic's signaling server, orchestrator, function store, database and web server in the \emph{us-east-2} region.
Additionally, multiple \magic peers are provisioned across 10 AWS regions\footnote{specifically, we use
{us-east-1}, {us-west-1}, {us-west-2}, {ca-central-1} , {sa-east-1}, {eu-central-1}, {eu-west-1}, {eu-west-2}, {eu-west-3} and {eu-north-1}}; again we use Node.js processes instead of browsers, also because AWS does not support SGX yet.
We install the two \magic functions on all peers and provision a special peer called \emph{measurer} with no functions installed in the \emph{us-west-1} region.
Note, that no clients share a region with the database server for simulating realistic latencies.
However, we intentionally let one regular peer and the measurer share the same region to represent nearby peers.
In this measurement, we compare three configurations:
\begin{inlinelist}
	\item 1 peer close to the measurer;
	\item 20 peers, \ie, 2 peers per region; and
	\item the baseline, where functions are invoked on server-side via HTTP.
\end{inlinelist}
We let the measurer issue invocation requests to \magic peers at fixed rates and measure the call latency, excluding connection establishments.

First, we let the measurer invoke the I/O intensive function \texttt{getGenres} at fixed throughputs from 1 to 1400 requests per second; in steps of 100.
We stop measuring, if the latency exceeds 1000\unit{ms} because this is a clear sign of the system being saturated.
\Cref{fig:tput-lat-both} shows the average call latency in dependence of the throughput for this function.
We see that invoking functions through \magic induces an up to $32\%$ higher latency due to an additional network hop between the measurer and the database server.
The configuration with one peer achieves stable latencies for up to $538$ requests per second, which is a $39\%$ lower throughput than the baseline. Here, the single peer cannot handle as many requests as the more powerful server machine and becomes a bottleneck.
When looking at the configuration with 20 peers, we see a higher throughput of up to $1.264$ requests per second, $1.67\times$ higher than the baseline.
The improvement is limited, as the bottleneck is now the centralised database server for this I/O intensive function.

Second, the measurer invokes the computationally intensive function \texttt{getRecommendations}.
We use the same requests rates as before, but add measurements between 1 and 25 requests per second.
\Cref{fig:tput-lat-both} shows the average call latency depending on the throughput.
Compared to the previous measurement, we report 
\begin{inlinelist}
	\item an overall lower throughput because the benchmark is CPU bound, and
	\item an overall higher latency, because the functions takes longer to complete.
\end{inlinelist}
Again,we see a higher latency due one additional network hop.
The configuration with 1 \magic peer performs similarly to the baseline, as both achieve up to 24 requests per second.
For more peers, \magic achieves up to 468 requests per second, which is $19.5\times$ the baseline and the single peer configuration; showing, that \magic scales linearly with increased number of peers.

To summarise, these two measurements show that \magic can achieve better scalability than a traditional system by offloading computation to peers.
We also see, that computation intensive functions are better candidates for offloading.
However, in contrast to the evaluation of the chat bot use case (see \Cref{sec:eval:lat-costs}), \magic peers experience higher latencies, because this use case contains a centralised database, resulting in an additional network hop.

\subsection{Unresponsive Peers and Reconnections} \label{sec:eval:reconn}
\begin{table}
\centering
\begin{tabular}{p{0.9cm}p{0.9cm}lllll}
\toprule
dwell time $d$ & RTT $r$ & \multicolumn{5}{c}{$P_{unresp.}$ for workload durations ($w$)}\\
& & 1\unit{ms} & 10\unit{ms} & 100\unit{ms} & 1\unit{s} & 10\unit{s}\\
\cmidrule(lr){3-7}

             &  10\unit{ms} & 0.01\% & 0.03\% & 0.18\% & 1.68\% & \cellcolor{black!25}16.68\%\\
 1\unit{min} & 100\unit{ms} & 0.09\% & 0.10\% & 0.25\% & 1.75\% & \cellcolor{black!25}16.75\%\\
             & 400\unit{ms} & 0.34\% & 0.35\% & 0.50\% & 2.00\% & \cellcolor{black!25}17.00\%\\
 \midrule
             &  10\unit{ms} & 0.00\% & 0.00\% & 0.01\% & 0.08\% & 0.83\%\\
20\unit{min} & 100\unit{ms} & 0.00\% & 0.01\% & 0.01\% & 0.09\% & 0.84\%\\
             & 400\unit{ms} & 0.02\% & 0.02\% & 0.03\% & 0.10\% & 0.85\%\\
 \midrule
             &  10\unit{ms} & 0.00\% & 0.00\% & 0.00\% & 0.04\% & 0.42\%\\
40\unit{min} & 100\unit{ms} & 0.00\% & 0.00\% & 0.01\% & 0.04\% & 0.42\%\\
             & 400\unit{ms} & 0.01\% & 0.01\% & 0.01\% & 0.05\% & 0.43\%\\
\bottomrule
\end{tabular}
\caption{Probabilities of unresponsive peers in dependence of different dwell times, round-trip times and workload processing durations.}
\label{tab:prob}
\end{table}

In a real-world \magic application, peers are expected to become unresponsive due to multiple reasons:
\begin{inlinelist}
	\item users leaving the websites (\eg closing their browser or specific browser tabs);
	\item external reasons such as network issues; or
	\item malicious peers that do not respond to invocation requests on purpose.
\end{inlinelist}
For the first class of events, we introduce a probability model to estimate the probability of peers being unresponsive $P_{unresp.}$, which is described in the following.
We exclude the second and third class from our model, as such events are unpredictable.
We consider two peers $A$ and $B$, while peer $A$ sends invocation requests to peer $B$.
We introduce the following variables:
$d$, the dwell time of peer $B$ (\ie, the time the user spends on the website), the WebRTC round-trip time $r$ between $A$ and $B$, and $w$, the duration the workload needs to finish processing.
Peer $B$ will be unresponsive, if the invocation request is sent in the last $l = w + \frac{r}{2}$ seconds of the dwell time $d$. 
Assuming a uniform distribution of function invocations, we can estimate the probability $P_{unresp.}$ of an invocation request remaining unanswered with
\[ P_{unresp.} = \frac{w+\frac{r}{2}}{d} \]
To calculate the probabilities, we assume the following realistic ranges for the introduced variables:
Users spend between 70\unit{seconds} and 40\unit{minutes} on website, depending on the type~\cite{liu2010understanding,schneider2009understanding}. 
Therefore, we assume 1, 20 and 40 minutes for the dwell time $d$.
For \webrtc round trip times, we use the values of 10\unit{ms}, 100\unit{ms} and 400\unit{ms} reported in~\cite{taheri2015webrtcbench} for local, university and mobile networks.
Since workload runtimes are application-specific, we assume values from 10\unit{ms} to 10\unit{s} for $w$. 
\Cref{tab:prob} shows the calculated probabilities for these ranges.
We see, that for the most combinations, probabilities are well below $0.1\%$ and only exceed $10\%$ when the workload processing time ranges near the peer's dwell time (shaded cells in \Cref{tab:prob}).
\magic's frequent \webrtc keep-alive messages further decrease these probabilities, as broken connections are detected earlier.
For \magic, this means that websites with expected shorter dwell times are more suited for shorter workload executions, while websites with expected longer dwell times (such as social networks) are more capable of handling longer workload execution times.

We use the reported probabilities for the following experiment:
We deploy a WebAssembly function performing an integer addition as a \emph{Cloudflare Worker}~(see \Cref{sec:impl}) and on \magic.
Choosing configured timeouts from 100 to 1,000\unit{ms}, we examine probabilities of 0.1\%, 1\% and 5\%.
These are induced by randomly dropping invocation requests with the given probability.
A fully responsive peer (\magic baseline) and the Cloudflare deployment (FaaS fallback) are both unaffected by the configured timeout and act as baselines.
\begin{figure}[t]
\includegraphics[width=\linewidth]{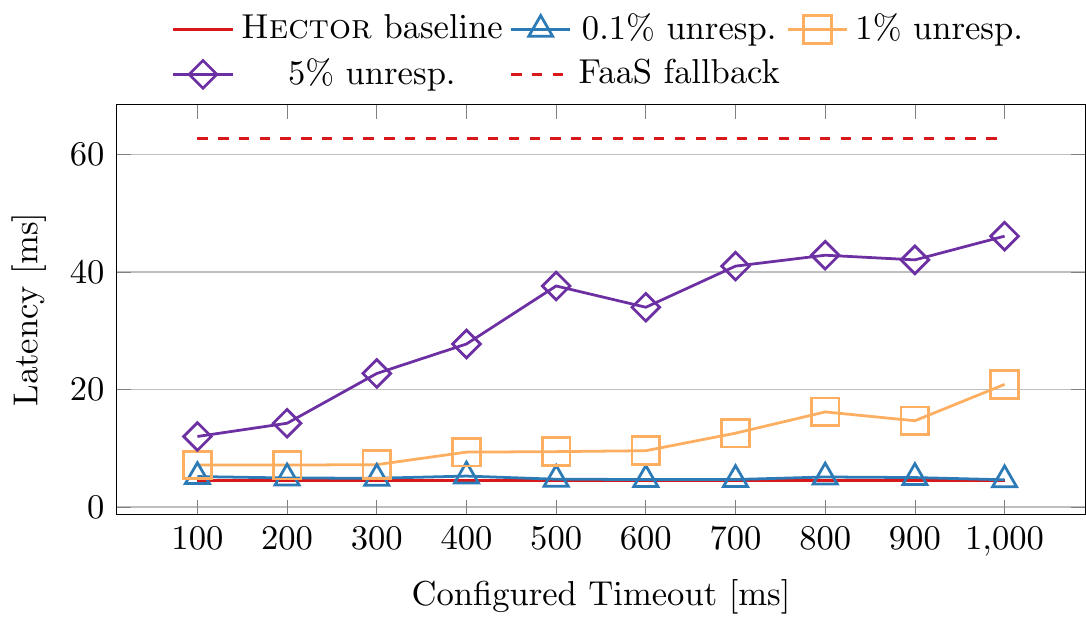}
\caption{Function invocation latencies for different shares of unresponsive peers and various timeouts.}
\label{fig:ignore-lat}
\end{figure}
We invoke every function 1000 times and report the average latency for a successful reply.
The results in \Cref{fig:ignore-lat} show, that the \magic baseline performs best with an average latency of 4.5\unit{ms}, closely followed by 0.1\% unresponsive peers with 4.9\unit{ms} average latency.
All invocations to partly unresponsive peers stay below the average latency of the FaaS fallback of 62.7\unit{ms}.
Our measured values only come close to this baseline, if 5\% of the peers are unresponsive and a relatively large timeout is configured.
Combining these insights with the ones from our probability model, we conclude that a \magic deployment will be largely unaffected by unresponsive peers.

\subsection{Security Analysis}\label{sec:attacks}
In the following, we discuss possible attacks against \magic according to the threat model (see \Cref{sec:bg:threatmodel}). 

\mypar{\ac{DoS} Attacks}
Local attackers can stop or deny starting enclaves at any time.
In typical SGX threat models~\cite{Schuster:2015:vc3,brenner2016securekeeper,arnautov2016scone,weiser2017sgxio,priebe2018enclavedb,goltzsche2018endbox}, \ac{DoS} attacks are excluded, as there are no countermeasures.
In \magic, such an attack only affects the attacker's machine and thus does not compromise the whole system.
Whenever a client sends a request to a peer and does not receive an answer within a timeout, it will retry a different peer and ultimately use the fallback.
Users that deny function execution can always use \magic's fallback infrastructure.
While such an action might reduce the service provider's cost savings, it is not security critical.
If the service provider applies some form of reimbursement for participating clients (see \Cref{sec:design:nutshell}), approaches like \cite{goltzsche2019acctee} can be used to verify that certain amounts of computations have actually been performed.

\mypar{Overloaded Peers}
An attacker can flood a peer with invocation requests aiming to overload that peer's processing power.
Whenever a \magic client observes an abnormal high number of requests from a single peer,
the client terminates the connection to that peer and blacklists it.
Therefore, an attacker is able to force a client to process unnecessary requests
but is not able to overload the machine.

\mypar{Unexpected Function Installation}
The only entry point to the enclave, besides processing a message from a peer,
is responsible for loading a function.
A \magic function is shipped in an encrypted file containing a WebAssembly 	or JavaScript code.
If decryption fails, the enclave raises an error.
Therefore, an attacker can instruct the enclave to install only functions encrypted and signed by the service provider.
Consequently, the attacker is able to install all available functions locally.
However, as this is invisible to the orchestrator, other clients will never connect to this peer.
\\ \\
All connections in \magic are encrypted with TLS.
However, as this connection is not terminated inside an SGX enclave, an additional security layer, as described in \Cref{sec:design:security}, is necessary.
This layer can protects against the following attacks.

\mypar{Modified or Forged Messages}
Due to the untrusted side of the system being responsible for accessing the network,
an attacker is able to drop, modify or forge any network packets.
All modified messages will be detected due to a wrong \ac{MAC} and dropped.
Message forging is not possible, as they are encrypted using the script secret (see \cref{sec:design:security}).

\mypar{Replayed, Dropped or Out-of-Order Messages}
In \magic, all responses are verified to contain the same nonce used in the request to prevent replay attacks.
Incorporating a monotonic counter in requests, enables the detection of messages being dropped or sent out-of-order.

\section{Related Work}
\label{sec:rw}

Distributing computations across personal machines of end-users is a well explored topic (\eg \cite{korpela2001seti,milojicic2002peer,anderson2004boinc,beberg2009folding}).
In this section, we focus on research closer related to \magic.

\mypar{Offloading to Browsers}
Several systems explore offloading computations or data to browsers, but most do not make use of trusted execution at client-side:
Akamai Netsession~\cite{zhao2013peer} is a commercially available \ac{CDN} that is capable of offloading traffic to participating peers without support for offloading computations.
Maygh~\cite{Zhang2013maygh} is a \ac{CDN} consisting of web browsers, offloading the delivery of static content to many clients.
Similarly to \magic, the system is based on WebRTC and includes one or more coordinators that manage the available peers.
The main difference to \magic is that Maygh can only deliver static content, but does not support the sharing of computation results, due to missing mechanisms to trust in them.
%
CloudPath~\cite{mortazavi2017cloudpath} enables offloading of server-side functionality to different locations on the path to the clients to reduce latency.
The approach of CloudPath is orthogonal to ours, \ie, \magic clients could be integrated into CloudPath to add an additional execution location closest to other clients.
Multiple systems~\cite{van2017legion,jannes2019edge,lavoei2019pando} also apply WebRTC to distribute computations across web browsers.
However, in contrast to \magic, protecting these from untrusted peers is not considered.


%

\mypar{Trusted Offloading to Browsers}
Other works also combine trusted execution technology with web browsers.
The \magic browser extension presented in this paper is based on ideas of our workshop paper \trustjs~\cite{goltzsche17trustjs}.
It represents an early prototype on our way towards \magic with very preliminary evaluation without any use cases.
We reimplemented it to be compatible to the  novel WebExtensions API~\cite{webextensions-api}, added support for \webrtc and replaced the slow JavaScript interpreter by the high-performance JavaScript and WebAsssembly runtime V8.
Fidelius~\cite{eskandarian2018fidelius} protects user inputs into web applications from a malicious operating system. 
Similarly to \trustjs, the system splits the code into a trusted and an untrusted part, each part being executed in a separate JavaScript engine.  
The system uses \ac{SGX} enclaves and single-board computers to create a trusted path from input to output devices.
However, it lacks any support for distribution over multiple browsers. 
The goal of Cyclosa~\cite{pires2018cyclosa} is to protect personal data of users of web search.
It uses a peer-to-peer network between browsers to send the user's query to the web search service.
However Cyclosa only supports one single application, which is web search.
While any other type of web application cannot be implemented with Cyclosa, \magic could be used to implement it (see \Cref{sec:design:applications}).
PrivaTube~\cite{da2019privatube} is a browser-based \ac{CDN} focused on video streams and uses SGX to establish trust in browsers.
However, it considers neither direct communication between browsers nor trusted computations.

\mypar{Trusted Offloading to Non-Web Clients}
Finally, a group of related research projects not related to the web also offload computations to untrusted clients, protecting these in \acp{TEE}.
ETTM~\cite{dixon2011ettm} and EndBox~\cite{goltzsche2018endbox} offload middlebox functionalities to clients and enforce this using trusted execution technology.
Vrancken et al.\cite{vrancken2019securely} offload computations to non-web clients of a peer-to-peer network and protect it with an SGX enclave. 
%
AccTEE~\cite{goltzsche2019acctee} performs resource accounting of offloaded computations on servers as well as on clients.


\section{Conclusion}
\label{sec:concl}
In this paper, we presented \magic, a novel web framework that allows service providers to freely distribute their web application over all its current users.
\magic establishes trust into browsers by applying trusted execution technology to offload function execution securely.
Functions are directly invoked on nearby browsers using peer-to-peer communication.
In our evaluation, we showed that \magic copes well with unresponsive peers, has lower latencies than traditional deployments without generating costs and scales linearly with increasing numbers of participants.


\bibliographystyle{ieeetr}
\bibliography{paper.bib,endbox.bib,acctee.bib,rfc.bib}



\end{document}